\documentclass[aps,prl,twocolumn,superscriptaddress]{revtex4-2}
\usepackage{epsfig}
\usepackage{graphicx}
\usepackage{amsfonts,amssymb}
\usepackage{amsmath}
\usepackage{color}
\usepackage{times}
\usepackage{flushend}
\usepackage{xcolor}
\usepackage{hyperref}

\hypersetup{
    colorlinks=true, 
    linkcolor=blue, 
    citecolor=blue, 
    filecolor=magenta, 
    urlcolor=blue, 
}

\begin{document}

\title{Photo-Induced Enhancement of Critical Temperature in a Phase Competing Spin-Fermion System}

\author{Sankha Subhra Bakshi}

\affiliation{Indian Institute of Science Education and Research-Kolkata, Mohanpur Campus, Kolkata 741246, India}

\date{\today}

\begin{abstract}
Ultrafast optical excitation is known to destabilize long-range order in correlated systems, yet experiments have also reported the emergence of metastable phases, in some cases with enhanced critical temperatures. The microscopic origin of such light-induced stabilization remains unresolved. Here we investigate this problem within a minimal spin–fermion framework: a double-exchange model at half filling, augmented by ferromagnetic superexchange on a square lattice. In equilibrium, at half-filling the ordering temperature is set by the competition between kinetic-energy–driven antiferromagnetism and superexchange-induced ferromagnetism. Using quantum Landau–Lifshitz–Gilbert–Brown dynamics for localized spins combined with mean-field evolution of itinerant electrons, we demonstrate a nonthermal mechanism for stabilizing ordered phases. Photoexcitation creates a long-lived nonequilibrium carrier population that resists thermalization and reshapes the low-energy landscape, converting kinetic-energy–driven antiferromagnetism into ferromagnetism and enhancing the critical temperature. While model-specific, our results reveal a general microscopic pathway by which light can tip the balance between competing orders, suggesting routes toward optically engineered magnetism, charge-density-wave order, and superconductivity.
\end{abstract}

\maketitle

\textit{Introduction.-} Ultrafast optical excitation in correlated electron systems has become a powerful means to manipulate quantum phases far from equilibrium. By driving carriers across Mott, Peierls, or interband gaps, light can suppress, recover, or reorganize long-range order \cite{OS1,OS2,OS3,OS4,OS5}. Beyond transient responses, optical driving can also reshape the underlying free-energy landscape, stabilizing hidden or metastable phases inaccessible in equilibrium \cite{PICO1,PICO2,PICO3,PICO4,PICO5,PICO6,PIOO1,PIOO2}. These advances naturally raise the question: can light enhance the stability of an ordered phase, effectively raising its transition temperature $T_C$, rather than merely disrupting it?  

A striking realization has been reported in the ferromagnetic Mott insulator YTiO$_3$, where terahertz (THz) pulses increase the ferromagnetic transition temperature by nearly a factor of three \cite{PIHT}. YTiO$_3$ lies close to the magnetic phase boundary of $R$TiO$_3$ ($R =$ La $\rightarrow$ Y), where small variations in ionic size tune the ground state from antiferromagnet (AF) to ferromagnet (F) \cite{YTO1,YTO2,YTO3}. The fragile ferromagnetic order in YTiO$_3$ is therefore highly sensitive to perturbations. Key observations include: (i) the ability of photoexcitation to shift $T_C$ upward or downward depending on frequency, (ii) ultrafast enhancement of the order parameter within $\sim 50$~ps, and (iii) persistence of the photoinduced state on nanosecond timescales. These features point to an optically driven stabilization of ferromagnetism in a regime where thermal order is marginal.  

The microscopic origin of this stabilization remains debated. In YTiO$_3$, low-frequency phonon excitation has been proposed to modify exchange pathways \cite{PIHT-theory}, while the long lifetimes of photoinduced states suggest a possible role for nonequilibrium carrier populations. This motivates a complementary perspective: can photoexcited carriers themselves tilt the balance between competing orders and reshape the free-energy landscape? To address this, we study a minimal spin–fermion Hamiltonian that captures the competition between antiferromagnetism, promoted by the spin–fermion coupling, and ferromagnetism, arising from a direct superexchange. This setup mimics the delicate balance in YTiO$_3$. Unlike the THz-driven case, we focus on high-energy pumping across the gap, which directly generates long-lived electronic populations. Our Hamiltonian is:
\begin{equation}
H =  \sum_{\langle ij \rangle, \sigma} -h c^\dagger_{i\sigma} c_{j\sigma} 
    - D \sum_{i} \mathbf{S}_i \cdot \hat{\mathbf{s}}_i
    - J_F \sum_{\langle ij \rangle} \mathbf{S}_i \cdot \mathbf{S}_j
    - \mu \sum_{i} \hat{n}_i,
\end{equation}
where $c^{\dagger}_{i\sigma}$ ($c_{i\sigma}$) creates (annihilates) an itinerant electron with spin $\sigma$ on site $i$. The first term describes nearest-neighbor hopping with amplitude $h$. The second term represents the local Hund’s coupling $D$ between itinerant electron spins $\mathbf{s}_i$ and classical localized core-spins $\mathbf{S}_i$. The third term is a ferromagnetic superexchange between localized spins with strength $J_F$. The final term fixes the filling through the chemical potential $\mu$. $\hat{n}_i$ is the density operator $\sum_{\sigma}c^{\dagger}_{i\sigma} c_{i\sigma}$.

Previous theoretical approaches to nonequilibrium ordering have ranged from phenomenological Landau–Ginzburg descriptions \cite{GL1,GL2,GL3} to numerically exact methods such as exact diagonalization \cite{ED1,ED2}, DMRG \cite{DMRG1,DMRG2}, and DMFT \cite{DMFT0,DMFT1,DMFT2,DMFT3,DMFT4,DMFT5,DMFT6}. These are limited either by system size, lack of spatial correlations, or computational cost. Hybrid schemes combining real-time electron dynamics with semiclassical bosonic modes have recently emerged as a scalable alternative \cite{MFD1,MFD2,MFD3,MFD4,MFD5}. We employ such an approach here, simulating coupled electron–spin dynamics using a quantum Landau–Lifshitz–Gilbert–Brown (QLLGB) framework \cite{sauri2020,LLGB1,LLGB2}.  

We restrict ourselves to the half-filled case. The equilibrium phase diagram (Fig.~1) shows that increasing \(J_F\) drives a transition from antiferromagnetic to ferromagnetic order, with a broadened crossover separating the two regimes. Near this phase boundary the critical temperature is strongly suppressed, rendering the system highly susceptible to perturbations. Focusing on this fragile regime, we show that a laser pulse resonant with the DOS gap creates nonequilibrium carriers that enhance ferromagnetic correlations and increase $T_C$, providing a microscopic pathway to light-stabilized order in correlated systems.

\begin{figure}[t]
\includegraphics[width=6cm,height=4.1cm]{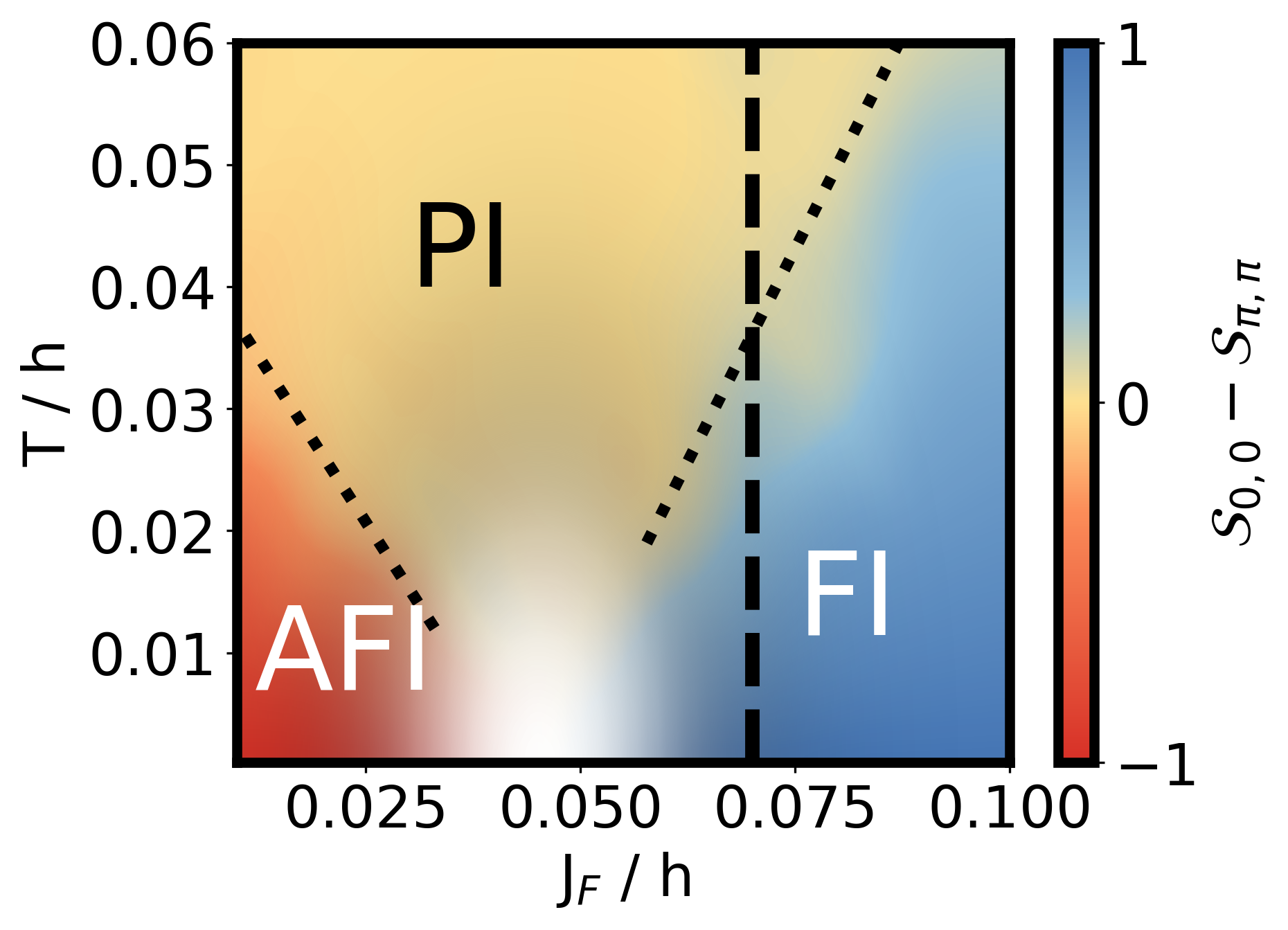}
\caption{
Equilibrium phase diagram of the model in Eq.~(1) in the \((T,J_F)\) plane at half-filling. Phase boundaries (dotted line) are obtained from the time-averaged structure factor evaluated at the ferromagnetic \((0,0)\) and antiferromagnetic \((\pi,\pi)\) ordering vectors. The color scale shows their difference, with red indicating an antiferromagnetic 
insulator (AFI), blue a ferromagnetic insulator (FI), and yellow colors a 
paramagnetic insulator (PI). For \(J_F/h < 0.03\), the system evolves from AFI to PI with increasing \(T\), while for \(J_F/h > 0.06\), the ground state is FM, crossing into PI at higher \(T\). The apparent fuzziness of the AF–FM boundary at low $T$ reflects broad momentum-space weight in the spin structure factor, originating from finite-size limitations in our simulations; in the thermodynamic limit this feature is expected to sharpen into a well-defined transition. The black dashed line marks our operating parameter point \(J_F/h=0.07\) for the rest of the study where the equilibrium critical temperature \(T_{C}/h \simeq 0.037\).
}
\end{figure}

\textit{Method.}-To incorporate the effect of an external laser pulse, we implement the Peierls substitution in the hopping term:
\(
-h~c^\dagger_{i\sigma} c_{j\sigma} \rightarrow -h~\exp\left[i \int_{\vec{R}_i}^{\vec{R}_j} \vec{A}(t) \cdot d\vec{R}\right]~c^\dagger_{i\sigma} c_{j\sigma},
\)
where \( \vec{R}_i \) denotes the position of site \( i \), and \( \vec{A}(t) \) is the time-dependent vector potential corresponding to an electric field \( \vec{E}(t) = -\partial \vec{A}(t)/\partial t \). The laser pulse is modeled using a Gaussian envelope centered at \( t = 0 \), with a full width of \( 30\tau_0 \) ($\tau_0 =1/h $) and carrier frequency resonant with the Hund’s coupling scale \( D \), facilitating interband excitation in the single-band model.

The classical core spins \( \mathbf{S}_i \) evolve via the stochastic Landau-Lifshitz-Gilbert-Brown (LLGB) equation:
\begin{equation}
\frac{d\mathbf{S}_i}{dt} = \mathbf{S}_i \times \left( \left\langle\frac{\partial H}{\partial \mathbf{S}_i} \right\rangle + \boldsymbol{\eta}_i \right) - \gamma~\mathbf{S}_i \times \left( \mathbf{S}_i \times \left\langle\frac{\partial H}{\partial \mathbf{S}_i} \right\rangle \right),
\end{equation}
where the first term represents precessional dynamics, the second accounts for Gilbert damping with damping coefficient \( \gamma \), and \( \boldsymbol{\eta}_i \) is a thermal noise term satisfying the fluctuation-dissipation theorem:  
\(\langle \boldsymbol{\eta}_i(t) \rangle = 0\), \(\langle \boldsymbol{\eta}_i(t) \boldsymbol{\eta}_j(t') \rangle = 2T\,\delta_{ij} \delta(t - t')\). The noise appears multiplicatively as \( \mathbf{S}_i \times \boldsymbol{\eta}_i \), preserving the norm \( |\mathbf{S}_i| =1\).

The electronic sector is described by the single-particle density matrix,  
\(
\rho_{j\sigma',i\sigma} = \langle c^\dagger_{i\sigma} c_{j\sigma'} \rangle,
\)
which evolves according to:
\(
\dot{\rho} = i\left[\rho, \mathcal{H}(\{\mathbf{S}_i\})\right],
\)
with the effective Hamiltonian given by
\(
\mathcal{H}_{i\sigma,j\sigma'}(\{\mathbf{S}_i\}) = -t_{ij} \delta_{\sigma\sigma'} - \frac{D}{2} \delta_{ij} \mathbf{S}_i \cdot \boldsymbol{\tau}_{\sigma\sigma'} -\mu \delta_{ij}\delta_{\sigma\sigma'},
\)
where \( \boldsymbol{\tau} \) are the Pauli matrices. Equation~(2) is integrated using a second-order Suzuki-Trotter decomposition \cite{Ma2011}, while Equation~(3) is evolved using a standard fourth-order Runge-Kutta (RK4) scheme. Unless otherwise specified, all simulations are performed on an \( 8\times8 \) square lattice with \( D/h = 20 \), a time step \( d\tau = \tau_0 / 100 \) and $\gamma=0.5$. 
\begin{figure}[t!]
\includegraphics[width=9cm,height=4cm]{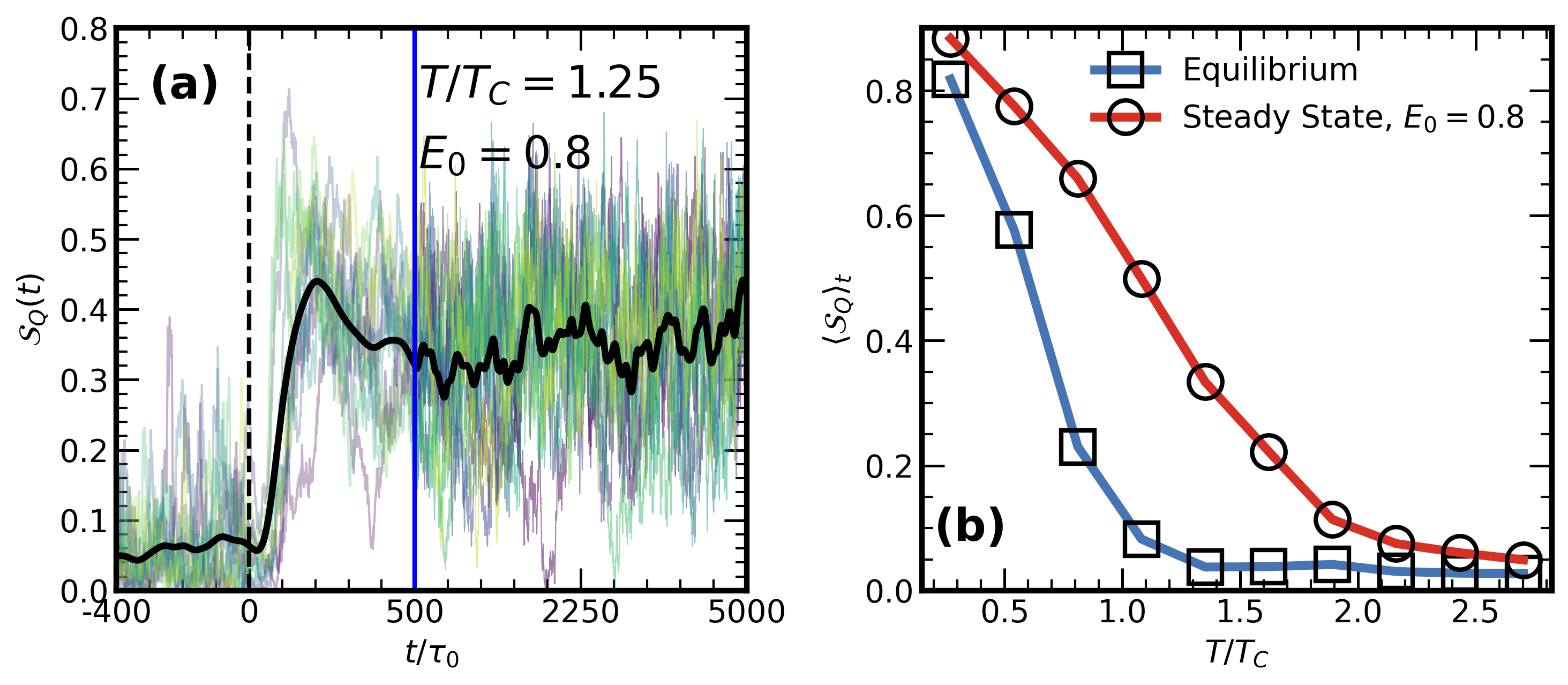}
\caption{
(a) Stochastic temporal evolutions of the ferromagnetic order parameter 
\(\mathcal{S}^{\alpha}_{Q}(t)\) for ten independent runs at bath temperature 
\(T/T_{C} = 1.25\) under a pump of strength \(E_0 \sim 0.8\). Different colors indicate different stochastic realizations \(\alpha\), while the black solid line denotes the run-averaged \(\mathcal{S}_{Q}(t)\). For \(t > 500\tau_0\), \(\mathcal{S}_{Q}(t)\) fluctuates 
around a steady-state value. (b) Steady-state value of \(\mathcal{S}_{Q}\), obtained by averaging over the interval \(1000\tau_0 \leq t \leq 5000\tau_0\) and over stochastic realizations, as a function of temperature \(T/T_{C}\). The equilibrium case (blue line) is compared with the nonequilibrium case under a pump of strength \(E_0 \sim 0.8\) (red line), which exhibits an approximate twofold enhancement of the critical temperature.  
}
\end{figure}

\textit{Enhancement of critical temperature.}-Strictly speaking, long-range magnetic order cannot exist in an ideal two-dimensional system with continuous symmetry due to the Mermin–Wagner theorem. In real materials, however, weak interlayer couplings, spin anisotropies, or other small symmetry-breaking terms lift this restriction, producing a sharp crossover that closely resembles a true phase transition. Our two-dimensional model should therefore be viewed as a minimal representation of such quasi-2D magnets: it captures the essential competition between ferro- and antiferromagnetism while remaining computationally tractable.  To characterize ferromagnetic order we monitor the spin structure factor at zero momentum,  
\(
\mathcal{S}_{Q} = \frac{1}{N^2} \sum_{i,j} \langle \mathbf{S}_i \cdot \mathbf{S}_j \rangle, \quad Q=(0,0),
\)  
which serves as an effective order parameter. Its stochastic evolution, obtained from independent Langevin trajectories at bath temperatures $T$, shows strong trajectory-to-trajectory fluctuations [Fig.~2(a)]. Nevertheless, following photoexcitation with pump strength $E_0\sim 0.8$, $\mathcal{S}_{Q}(t)$ fluctuates around a finite steady-state value. Averaging over long time windows and multiple stochastic realizations yields the nonequilibrium steady-state structure factor, $\langle \mathcal{S}_{Q}\rangle_t$, as a function of temperature [Fig.~2(b)]. Comparing equilibrium and photo-driven cases, we find that in equilibrium $\langle \mathcal{S}_{Q}\rangle_t$ collapses near \(T \simeq T_C\), while under pumping (\(E_0 \sim 0.8\)) correlations persist up to much higher temperatures. This corresponds to nearly a twofold enhancement of the apparent steady-state critical temperature, $T_{C}^{\mathrm{ss}} \simeq 2T_C$. The enhancement is found to be robust against variations in the phenomenological dissipation rate $\gamma$, as detailed in the Supplementary Material \cite{supp}. To understand the microscopic origin of this enhanced stability, we next examine the electronic sector.

\begin{figure}[t]
\includegraphics[width=8.5cm,height=4cm]{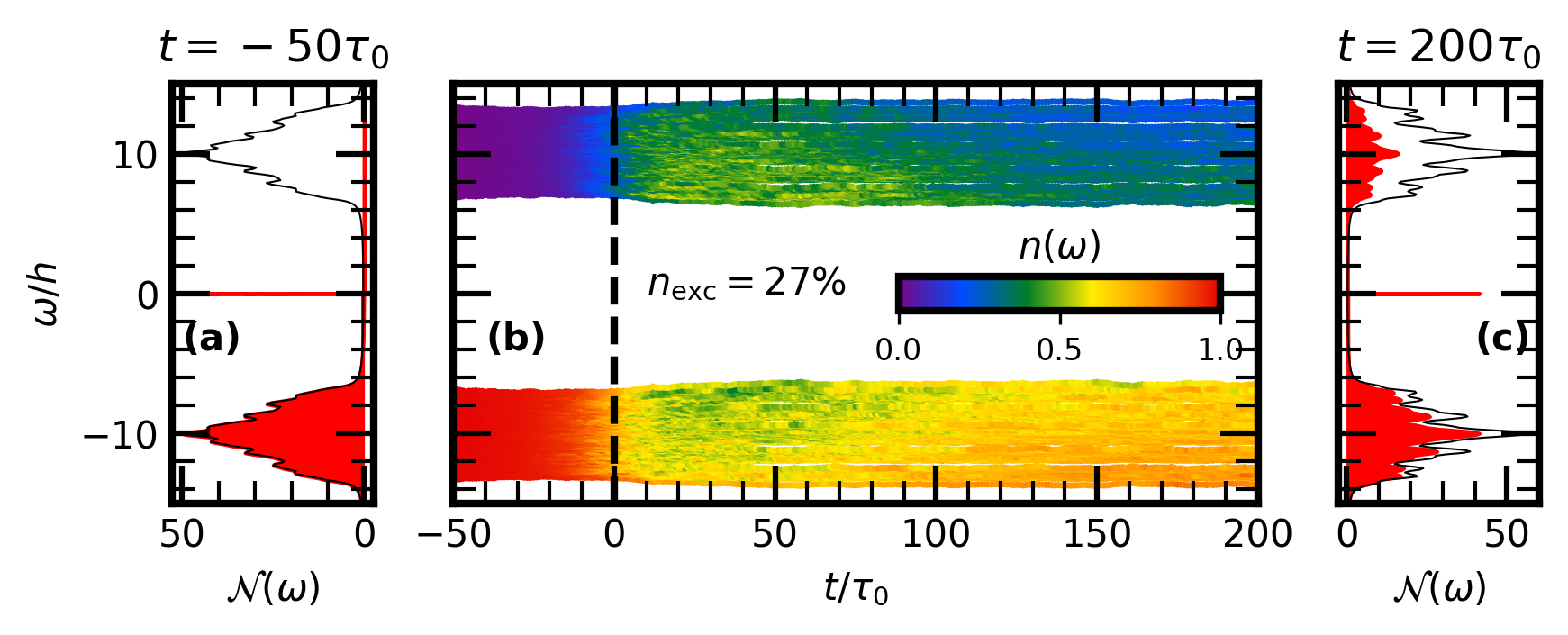}
\includegraphics[width=8.3cm,height=6cm]{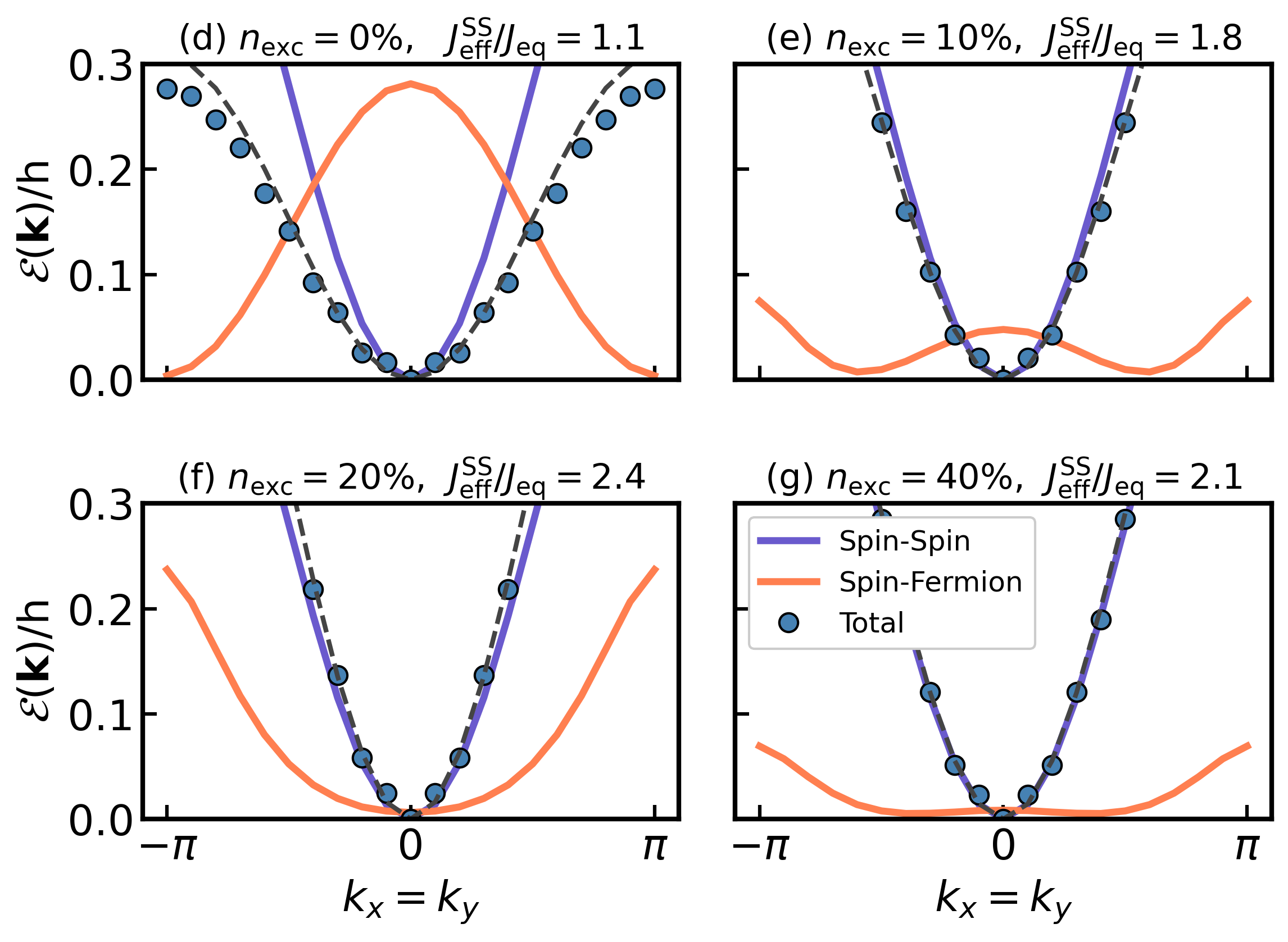}
\caption{
Top panels (a–c): Instantaneous electronic eigenvalues \(\epsilon_{\nu}\) and their occupations for pump strength \(E_0 = 0.8\), yielding an excited population \(n_{\mathrm{exc}} \approx 27\%\).  
(a) and (c) show the density of states (DOS, black) and occupied DOS (red shaded) before the pump (\(t = -50\tau_0\)) and after it (\(t = 200\tau_0\)); the pump width is \(30\tau_0\).  
(b) displays individual eigenenergies color-coded by occupation.  
Bottom panels (d–g): Potential energy surfaces \(\mathcal{E}(\mathbf{k}) = \mathcal{E}(\mathbf{k}) - \min[\mathcal{E}(\mathbf{k})]\) along \(k_x=k_y\) for excited populations \(n_{\mathrm{exc}} = 0.0, 0.1, 0.2, 0.4\).  
Violet: superexchange \(\mathcal{E}_{SE}\) (independent of \(n_{\mathrm{exc}}\)); red: electronic contribution \(\mathcal{E}_{el}\); green: total \(\mathcal{E} = \mathcal{E}_{SE} + \mathcal{E}_{el}\).  
In equilibrium (d), \(\mathcal{E}_{SE}\) favors ferromagnetism (F) while \(\mathcal{E}_{el}\) favors antiferromagnetism (AF).  
At \(n_{\mathrm{exc}}=0.1\) (e), \(\mathcal{E}_{el}\) weakens at \(\textbf{k}=\textbf{Q}\); at \(n_{\mathrm{exc}}=0.2\) (f), it flips to favor F, enhancing \(T_{C}\).  
At \(n_{\mathrm{exc}}=0.4\) (g), \(\mathcal{E}_{el}\) nearly vanishes and \(T_{C}\) approaches the bare superexchange limit.  
The dispersion \(\mathcal{E}(\mathbf{k})\) is fit with an effective exchange \(J^{\text{ss}}_{\text{eff}}\).  
}
\end{figure}
\textit{Electronic quasi-steady state.}-In the parameter regime studied, the spin timescale \(\tau_S \sim 1/J^{\mathrm{eq}}_{\mathrm{eff}}\), with $J^{\mathrm{eq}}_{\mathrm{eff}}$ the effective spin–spin interaction strength, is much longer than the electronic timescale \(\tau_0 \sim 1/h\) (by roughly two orders of magnitude), ensuring the hierarchy \(\tau_0 \ll \tau_S\). Exploiting this adiabatic separation, we evaluate the instantaneous electronic density of states (DOS) from the eigenvalues \(\epsilon_n(t)\) of the electronic Hamiltonian in the frozen spin background \(\{\mathbf{S}_i\}_t\). The corresponding occupations are obtained from the electronic density matrix \(\rho_{i\sigma,j\sigma'}(t)\). To characterize the electronic distribution, we define
\(
n(\omega,t)\,\mathcal{N}(\omega,t) = \sum_n \rho_{n}(t)\, \delta(\omega - \epsilon_n(t)), \quad
\mathcal{N}(\omega,t) = \tfrac{1}{N} \sum_n \delta(\omega - \epsilon_n(t)),
\)
where \(\rho_{n}(t)\) denotes the occupation of the \(n\)-th instantaneous eigenstate with energy \(\epsilon_n(t)\), obtained as
\(
\rho_{n}(t) = \sum_{ij, \sigma\sigma'} 
\Gamma^*_{i\sigma, n}(t)\, \Gamma_{j\sigma', n}(t)\, \rho_{ij}^{\sigma\sigma'}(t),
\)
with \(\Gamma(t)\) the instantaneous eigenvectors.  
The top panels of Fig.~3 illustrate the effect of the laser pulse on the electronic spectrum. Panels (a) and (c) show the DOS (black lines) before and after the quasi-steady state is reached. The DOS itself remains essentially unchanged, with no significant gap renormalization. In contrast, the occupied DOS, \(n(\omega,t)\mathcal{N}(\omega,t)\) (red shading), exhibits a marked redistribution: population is transferred to the upper band, accompanied by hole creation in the lower band. For pump strength \(E_0 \sim 0.8\) at $T=1.25~T_C$, the excited fraction \(n_{\mathrm{exc}}\) is about 27\% and remains stable for timescales at least up to \(5000 \tau_0\). A detailed analysis of \(n_{\mathrm{exc}}\) as a function of \(E_0\) is presented in Ref.~\cite{supp}.  
The persistence of these nonequilibrium populations, despite coupling to an external bath, signals a thermalization bottleneck analogous to that in Mott insulators \cite{therm1,therm2}. In our case, relaxation via particle–hole recombination requires the emission of multiple bosons, since the electronic excitation scale \(\sim D\) is much larger than the characteristic magnetic scale \(J_{\mathrm{eff}}\) and the bath bandwidth \(\Delta_{\mathrm{bath}}\). The corresponding thermalization time is therefore exponentially enhanced,
\(
\tau_{\mathrm{therm}} \sim \tau_0 \, e^{\lambda (D/\Delta_{\mathrm{bath}})},
\)
with \(\lambda \sim \mathcal{O}(1)\), leading to decay times far beyond our simulation window. This explains the long-lived nonequilibrium carrier populations that define the quasi-steady state in our calculations.
This redistribution of occupations provides the basis for constructing the nonequilibrium potential energy surfaces shown in Figs. 3(d–g).

\textit{Stiffness from nonequilibrium potential energy surface.}-In the presence of a long-lived upper-band population \(n_{\mathrm{exc}}\), the excitation can be treated as a conserved parameter and its distribution can be effectively described by a Fermi function with effective electronic temperature $T_{e}$ \cite{supp}. This allows us to construct a nonequilibrium potential energy surface in momentum space, characterizing the energy per site for spin configurations with ordering vector \(\mathbf{k}\):  
\(
\mathcal{E}(\mathbf{k}) = \mathcal{E}_{el}(\mathbf{k})\ + \mathcal{E}_{SE}(\mathbf{k}).
\)
The first term, \(\mathcal{E}_{el}(\mathbf{k})=\frac{1}{N} \int_{-\infty}^{\infty} \omega\, d\omega\, \mathcal{N}(\omega,\{\mathbf{S}_{\mathbf{k}}\})\, n(\omega) \), arises from the electron–spin coupling and depends on the electronic occupation \(n(\omega)\), while the second term, \(\mathcal{E}_{SE}(\mathbf{k}) =  -2J_F (\cos k_x + \cos k_y)\), represents the bare superexchange contribution, independent of electronic dynamics.  
Figures~3(d)–(g) show \(\mathcal{E}(\mathbf{k})\) along the diagonal path \(k_x=k_y\) for increasing upper-band populations \(n_{\mathrm{exc}}=0.0,0.1,0.2,0.4\). In equilibrium [Fig.~3(d)], the contributions compete: \(\mathcal{E}_{SE}\) favors ferromagnetic (F) order, while \(\mathcal{E}_{el}\) promotes antiferromagnetic (AF) correlations. As \(n_{\mathrm{exc}}\) grows, the electronic contribution first weakens [Fig.~3(e)], then reverses sign to support ferromagnetism [Fig.~3(f)], and eventually becomes negligible [Fig.~3(g)].  
By fitting the total dispersion to an effective Heisenberg form, we extract the nonequilibrium spin stiffness \(J^{\mathrm{neq}}_{\mathrm{eff}}\), which displays a nonmonotonic dependence on \(n_{\mathrm{exc}}\), peaking near \(n_{\mathrm{exc}}\sim 0.2\). Importantly, this maximum exceeds the bare superexchange scale \(J_F\), demonstrating that the enhancement of magnetic stiffness originates from an active electronic mechanism rather than a simple suppression of the AF contribution.

\begin{figure}[t]
\includegraphics[width=5cm,height=4cm]{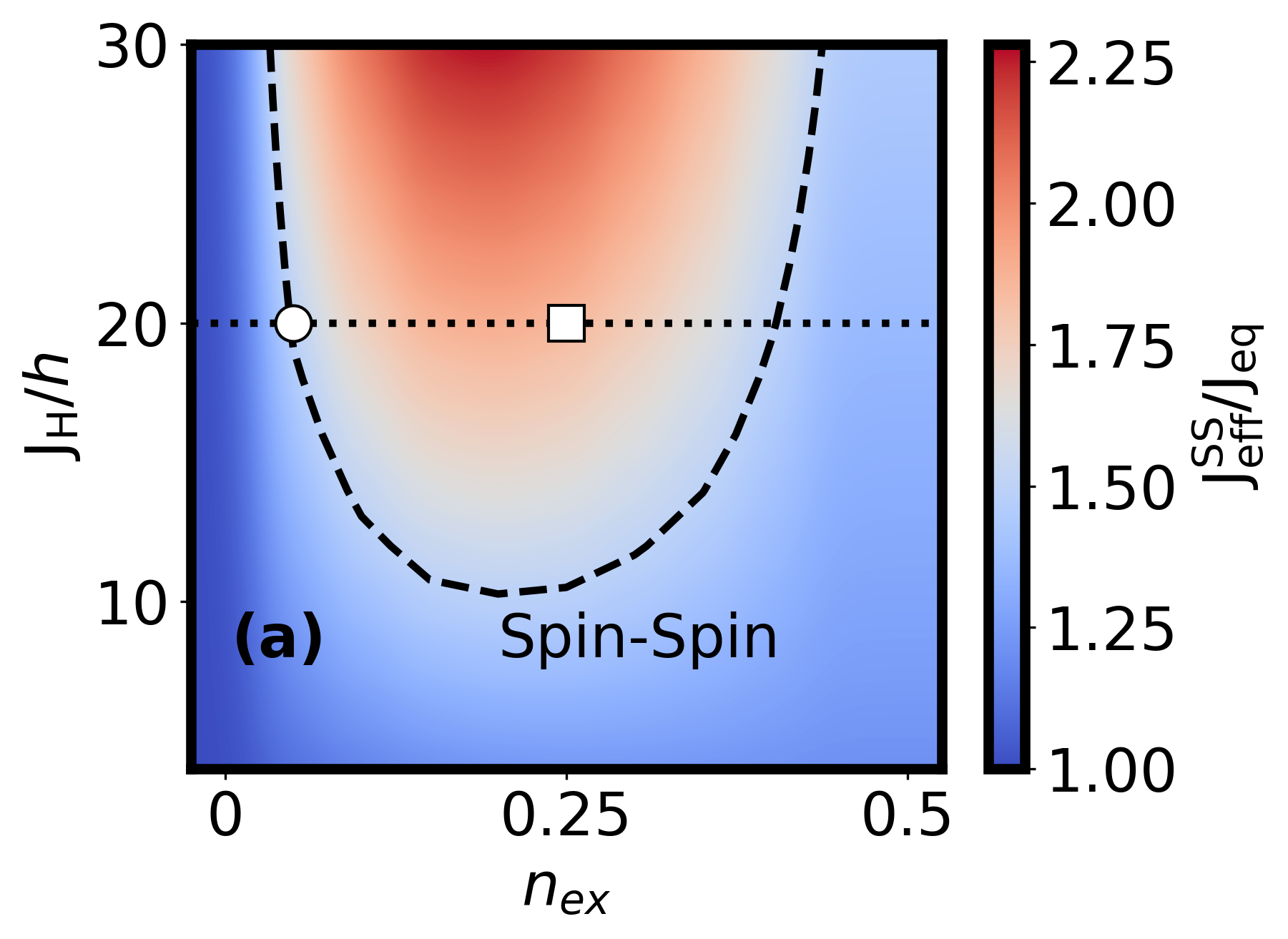}
\includegraphics[width=3.5cm,height=3.9cm]{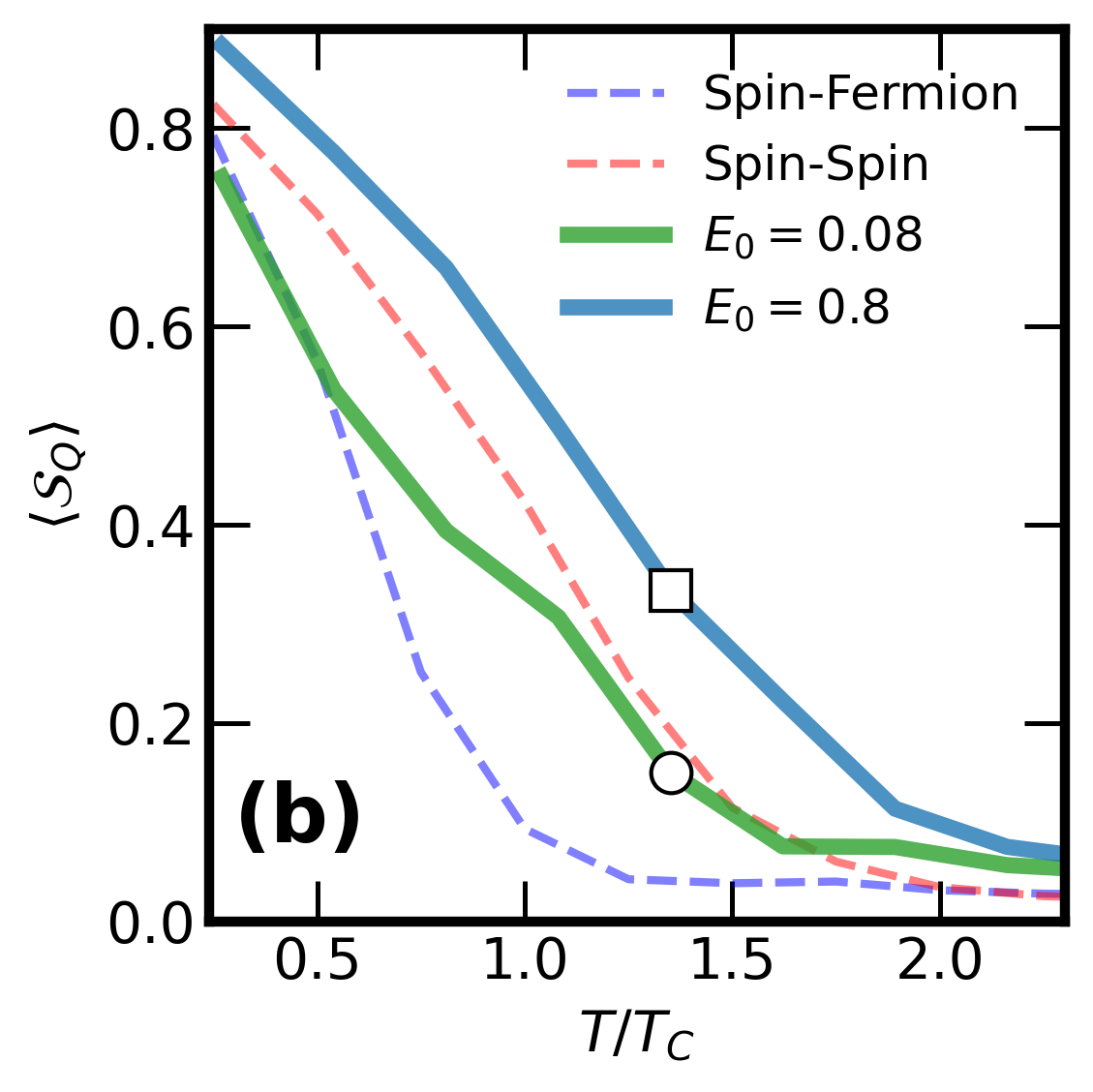}
\caption{
(a) Normalized effective exchange interaction, \(J^{\text{ss}}_{\text{eff}}/J_{\text{eq}}\), 
as a function of Hund’s coupling \(D/h\) and excited carrier density 
\(n_{\text{exc}}\) plotted as a color-map. The dashed black curve denotes the ferromagnetic exchange 
scale \(J_F\), which sets the upper bound of the effective stiffness in the 
fully nonequilibrium limit where the antiferromagnetic electronic contribution 
vanishes. The maximum enhancement of \(J^{\text{ss}}_{\text{eff}}\) arises in the 
intermediate nonequilibrium regime, indicating that while the equilibrium 
spin–fermion exchange is antiferromagnetic, suitable electronic population 
redistribution can drive it ferromagnetic.  
(b) Order parameter \(\langle S_{Q}\rangle\) as a function of temperature \(T\) for two different pump strengths $E_0$. 
The equilibrium result is shown by the blue dotted line, while the red dotted 
line indicates the bare spin–spin interaction. The green and blue curves 
correspond to pump strengths \(E_0 = 0.08\) and \(E_0 = 0.8\), respectively. 
At \(T/T_{C} = 1.25\), \(E_0 = 0.08\) produces an excited carrier density 
\(n_{\text{exc}} \approx 11\%\) (circle), while \(E_0 = 0.8\) yields 
\(n_{\text{exc}} \approx 27\%\) (square). The corresponding \(J^{\text{ss}}_{\text{eff}}\) 
values are marked on the right axis using the same symbols.  
}

\end{figure}
We now address the central question: \textit{under what conditions is the enhancement of the critical temperature maximized?} The effective magnetic stiffness is determined by the competition between the bare ferromagnetic exchange \(J_F\) and the electron-mediated antiferromagnetic exchange \(J_{\text{ex}}\):  
\(
J_{\text{eq}}= J_F - J_{\text{ex}}.
\)
In equilibrium and in the limit \(D/h \gg 1\), the antiferromagnetic contribution scales as \(J_{\text{ex}} \sim h^2/D\), opposing the bare ferromagnetic term. Nonequilibrium excitation qualitatively modifies this balance: as the excited carrier density approaches spin degeneracy (\(n_{\text{exc}} \sim 0.5\)), \(J_{\text{ex}}\) vanishes and the stiffness saturates at the bare value \(J_F\). At intermediate excitation, \(J_{\text{ex}}\) changes sign, leading to a regime where the effective stiffness exceeds its equilibrium value before being suppressed again at high excitation.  
To explore this systematically, we parameterize \(J_F = (1+\zeta)\, h^2/D\), such that in equilibrium the stiffness reduces to \(J \simeq \zeta h^2/D\). Choosing \(\zeta = 1/2\), we vary both \(D\) and \(n_{\text{exc}}\). The resulting phase diagram [Fig.~4(a)] shows the normalized stiffness \(J^{\text{ss}}_{\text{eff}}/J_{\text{eq}}\) as a function of \(D/h\) and \(n_{\text{exc}}\). The dashed curve indicates the bare super-exchange bound \(J^{\text{ss}}_{\text{eff}}=J_F\), reached when \(J_{\text{ex}}\to 0\). The maximum enhancement occurs in the intermediate regime, demonstrating that photoexcitation does more than simply melt electronic order: it actively reshapes the exchange pathways by redistributing carrier populations.  
This effect is directly reflected in the magnetic order parameter [Fig.~4(b)]. At moderate excitation (\(E_0=0.08\), corresponding to \(n_{\text{exc}} \approx 11\%\)), the order parameter is enhanced relative to equilibrium, consistent with the increase of \(J^{\text{ss}}_{\text{eff}}\). At stronger pumping (\(E_0=0.8\), \(n_{\text{exc}} \approx 27\%\)), the enhancement is even more pronounced, with the extracted stiffness values aligning quantitatively with the phase diagram in Fig.~4(a).

\textit{Discussion.-} Our results differ from the experimental observations of Ref.~\cite{PIHT} in several important respects. First, we employ a single-band model in which nonequilibrium carrier populations are generated by a high-frequency pulse with energy comparable to the band gap ($\sim D$), whereas the experiment uses a low-frequency pump resonant with phonon modes. Given the multi-orbital character of the experimental system, phonon-driven exchange modification is a plausible route; concurrently, tunneling and interband processes could produce long-lived carriers. Our study therefore explores the latter scenario as a complementary possibility rather than a literal microscopic description of the experiment.  
Second, in our model the ferromagnetic superexchange is treated as a classical, state-independent contribution; photoexcitation primarily suppresses the electron-mediated antiferromagnetic channel without directly renormalizing $J_F$. Finally, we observe no appreciable carrier relaxation up to $\sim 5000\tau_0$. While a more refined treatment should incorporate orbital degrees of freedom and beyond–mean-field fermionic correlations, the methodology developed here-notably the explicit calculation of nonequilibrium electronic stiffness-provides a concrete framework to assess how carrier redistribution alters exchange pathways. This perspective complements phonon-driven scenarios and suggests experimentally testable signatures that can distinguish the dominant microscopic route in specific materials.

\textit{Conclusion.-} We have demonstrated that in a spin–fermion model with competing antiferromagnetic (from spin–fermion coupling) and ferromagnetic (from superexchange) interactions, photoexcitation can substantially enhance the magnetic ordering temperature. A laser pulse resonant with the density-of-states gap generates a long-lived nonthermal carrier population that dynamically renormalizes spin interactions to favor ferromagnetism. Consequently, the effective magnetic stiffness \(J^{\text{ss}}_{\text{eff}}\) exceeds its equilibrium value, yielding a robust enhancement of the critical temperature.  
Our work establishes a minimal and general framework for how nonequilibrium carrier populations can stabilize ordered phases, with implications beyond ferromagnetism. While the present QLLGB approach is tailored to spin–fermion systems, the combination of nonequilibrium potential-energy surfaces with real-space stochastic dynamics provides a general methodological template that can be extended to more complex photoexcited correlated systems, including Mott insulators, electron–phonon coupled materials, and superconductors \cite{PISC}.

\vspace{1cm}
\textit{Acknowledgment:} Author would like to acknowledge financial support from the Indo-French Centre for the Promotion of Advanced Research (CEFIPRA). Author would also like to thank Prof. Pinaki Majumdar for discussions and valuable suggestions.

\clearpage
\newpage
\onecolumngrid
\setcounter{figure}{0}

\begin{center}
\fontsize{12pt}{16pt}\selectfont
\textbf{Supplementary to `Photo-Induced Enhancement of Critical Temperature in a Phase Competing Spin-Fermion System'} \\[1em]

\fontsize{11pt}{14pt}\selectfont
Sankha Subhra Bakshi \\[0.5em]
\textit{Indian Institute of Science Education and Research-Kolkata, Mohanpur Campus, Kolkata 741246, India}
\end{center}

\vspace{1em}

\section{Dependence on dissipation rate $\gamma$}
The dissipation rate $\gamma$ is treated as a phenomenological constant in our analysis. Throughout the main text we fix $\gamma = 0.5$. In Fig.~1. we demonstrate that the enhancement of order does not depend sensitively on the precise value of $\gamma$: varying $\gamma$ in the range $0.1$--$1.1$ changes the steady-state value by only about $\sim 10\%$, for a fixed pump strength $E_0 = 0.8$ and bath temperature $T = 1.25\,T_{C} > T_{C}$. While the transient dynamics can vary with $\gamma$, our focus here is restricted to the steady state.  
\begin{figure}[h]
\includegraphics[width=13cm,height=4cm]{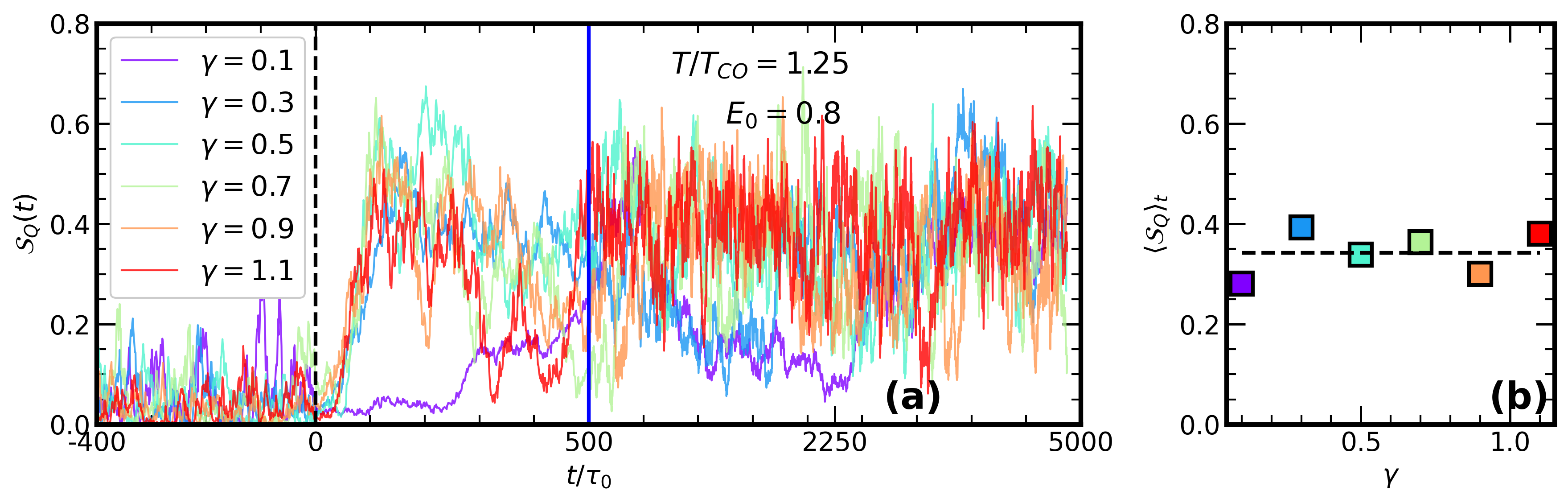}
\caption{
(a): Time evolution of the ferromagnetic structure factor \(S_Q(t)\) at pump 
strength \(E_0 = 0.8\) and bath temperature \(T/T_{C} = 1.25\) for dissipation 
factors \(\gamma\) ranging from 0.1 to 1.1.  
(b): Steady-state time-averaged structure factor \(\langle S_Q \rangle_t\) as a 
function of \(\gamma\), showing similar values across this range with variations 
within \(\sim 10\%\).  
}
\end{figure}
\section{Nonequilibrium population}
Fig.~2 shows how the nonequilibrium population responds to a pump. In (a), the occupations of instantaneous eigenstates are compared before (blue) and after (red) the pump, with a Fermi fit at an effective temperature \(T_e \sim 0.6D\). Panel (b) tracks the time evolution of the upper-band population \(n_{\mathrm{exc}}(t)\) for different pump strengths \(E_0\), highlighting the transient response. Panel (c) presents the steady-state upper-band population as a function of \(E_0\).
\begin{figure}[h]
\includegraphics[width=13cm,height=4cm]{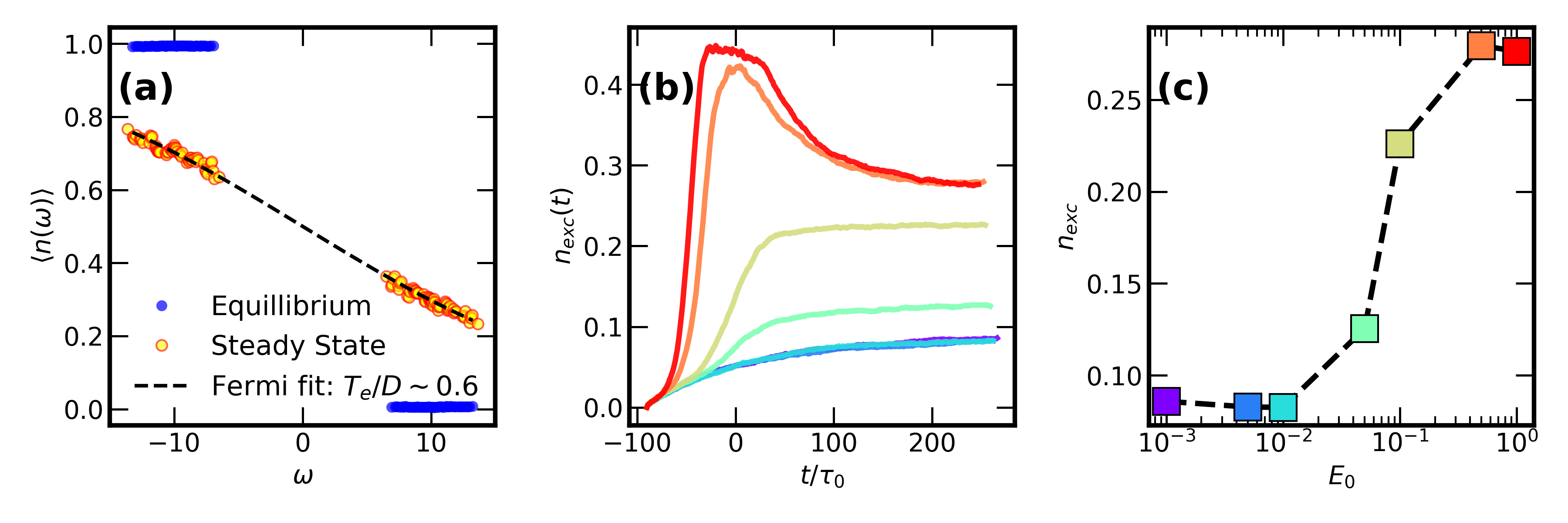}
\caption{
Nonequilibrium population dynamics at bath temperature \(T/T_{C} = 1.25\).  
(a) Occupations of instantaneous eigenstates in the equilibrium state before the 
pump (blue) and in the post-pump steady state (red). The black dotted line is a 
Fermi-function fit with electronic temperature \(T_e \sim 0.6D\).  
(b) Time evolution of the upper-band population \(n_{\mathrm{exc}}(t)\) for 
different pump strengths \(E_0\).  
(c) Steady-state values \(n_{\mathrm{exc}}(E_0)\) as a function of pump strength, 
with colors matching the trajectories in (b).  
}
\end{figure}

\end{document}